# Spatial polarization independent parametric upconversion of vectorially structured light

Hai-Jun Wu (吴海俊),[1] Bo Zhao（赵波）,[1] Carmelo Rosales-Guzmán,[1] Wei Gao (高玮),[1] Bao-Sen Shi (史保森),[1,2] and Zhi-Han Zhu (朱智涵) [1,*]

[1] *Wang Da-Heng Collaborative Innovation Center, Heilongjiang Provincial Key Laboratory of Quantum Manipulation & Control, Harbin University of Science and Technology, Harbin 150080, China*

[2] *CAS Key Laboratory of Quantum Information, University of Science and Technology of China, Hefei, 230026, China*

Spatial polarization independent (SPI) parametric conversion is the basis of many optical applications, such as SPI frequency interface for communication channels carried by vector modes and upconversion detection for polarization-resolved imaging. However, realizing such conversion remains a challenge. In this proof-of-principle work, we demonstrated SPI parametric upconversion using a polarization Sagnac nonlinear interferometer based on type-II second-harmonic generation (SHG). Our results show that the vector (including both polarization and intensity) profile and associated SOC state of the vector signal beam could be transferred to the SHG beam with a high fidelity. The principle lays a foundation of SPI frequency interface for quantum/classical channels based on vector modes and also paves the way for upconversion detection of polarization-resolved imaging in Mid-/far-infrared region.

*Introduction.* — The spin nature of a light field is determined by its state of polarization (SoP), and the field is commonly known as vectorially structured light field (or vector light for short) when the SoP is spatially non-uniform [1–3]. The feature of carrying a spatially-variant SoP originates from intrinsic spin-orbit coupling (SOC) within the light field [4, 5], hence the vector light can be regarded as a non-separable state of spin and spatial mode [6, 7]. Historically, vector light has been discovered soon after the invention of the laser [8, 9]; however, it takes several decades for researchers to reveal its unique focusing properties and the SOC nature. Thereafter, it has been widely studied and resulted in a number of innovations throughout the main fields of modern optics, such as imaging, optical trapping, communication and fundamental physics [10–19].

In this new subfield of modern optics, on-demand shaping and control of vector light is crucial for both fundamental and applied studies. For this task, a straightforward approach is manipulating the SOC structure of vector light via Geometric-phase elements with artificial microstructures [20]. Additionally, another promising approach is using nonlinear optical interactions; here, one can simultaneously manipulate the multiple degrees of freedom (DoFs) of light fields, including the frequency, spatial distribution of amplitude and SoP. Meanwhile, the SOC-mediated nonlinear polarization (NP) afforded by vector light, as an additional auxiliary interface, can significantly enhance our ability to shape and control nonlinear interactions. For this, recently, this topic has raised considerable interest in the community of nonlinear optics, ranging from parametric up/down-conversion (PDC) and stimulated Brillouin/Raman scattering to high-order harmonic generation [21–31].

Among the various nonlinear applications, optical frequency conversion based on parametric processes, such as sum-frequency generation and multi-wave mixing, is the most basic but also the most important one presently used in laser and photonic systems. Such a basic function, however, is not easy to accomplish when it meets a vector light or a polarization-resolved image. Because most optical parametric processes are sensitive to the SoP of input signal, while frequency conversion for vector light or polarization-resolved images requires spatial polarization independent (SPI) parametric conversion. The core of SPI parametric conversion is the ability to convert orthogonally-polarized components (spatial amplitudes) of input signal and keep their relative phase simultaneously. This ability, however, poses an experimental challenge, i.e. how to realize SPI parametric conversion by using currently available nonlinear crystals?

A feasible solution for this challenge would be the use of nonlinear interferometers with SU(2) symmetry [32], for which the two-crystal and Sagnac schemes (both are phase-locking structure) are two reliable configurations that have already become standard methods for the generation of polarization entanglement [33–35]. In particular, recently André G. de Oliveira *et al.* have realized real-time phase conjugation for vector light using SPI down conversion based on the two-crystal scheme [36]. Additionally, several groups have recently demonstrated the second-harmonic generation (SHG) of vector light via a two-crystal scheme consisting of orthogonal thin type-I crystals [37, 38], and we have also demonstrated this based on the Sagnac scheme with a thin type-I crystal, as well as a long type-0 crystal [39, 40]. Despite these works realized the frequency doubling of vector light, the SOC structure of input signal was also transformed. Namely, the SPI frequency conversion is still unrealized.

---

[*] zhuzhihan@hrbust.edu.cn

In this work, we report the SPI parametric upconversion based on the SU(2) nonlinear interferometer. Notably, while long periodically poled crystals are high efficiency, they are not easy to devise in the simple two-crystal scheme. Because cascading two long crystals will introduce a large distance between the two generation planes of the orthogonally-polarized components, leading to different diffraction progress for the two components. In view of this, here, a polarization Sagnac nonlinear interferometer with a type-II SHG crystal was adopted as the SPI frequency converter. Special attention is given to the fidelity of the converting apparatus, specifically, analyzing and comparing the SOC structure of the input signals and the corresponding output SHG.

*Methods and Results.* — General vector light can be regarded as a non-separable superposition with respect to a pair of orthogonal SoPs $\hat{\mathbf{e}}_\pm$ and associated SoP-dependent spatial modes. Here, without a loss of generality, both in theoretical analysis and experimental demonstration, we choose pairs of conjugate Laguerre-Gaussian (LG) modes as the SoP-dependent spatial modes; given by $LG^p_{\pm\ell}(r,\varphi,z) = u^p_{|\ell|}(r,\varphi,z)\exp[-i(kz\pm\ell\varphi)]$ in the cylindrical coordinates, where $\ell$ and $p$ denote the azimuthal and radial (spatial) indices of the LG modes, respectively, and $u^p_{|\ell|}(r,\varphi,z)$ is the envelop of the spatial amplitude. The corresponding vector light can be expressed as:

$$\mathbf{E}(r,\varphi,z) = \sqrt{\alpha}LG^p_{+\ell}(r,\varphi,z)\hat{\mathbf{e}}_+ + e^{i\theta}\sqrt{1-\alpha}LG^p_{-\ell}(r,\varphi,z)\hat{\mathbf{e}}_-, \quad (1)$$

where $\alpha \in [0,1]$ and $\theta$ are the weight coefficient and intermodal phase, respectively. It should be noted that $LG^p_{\pm\ell}(r,\varphi,z)$ carry the opposite orbital angular momentum (OAM), i.e., $\pm\ell\hbar$ per photon, but have the same order number of spatial DoF, given by $N = 2p + |\ell|$ [41]. This indicates that each pair of conjugate LG modes spans a spin-like SU(2) space with a mode order of $N$. In particular, Ref. 42 showed that the corresponding geometric description can be regarded as an OAM equivalent of the SoP Poincaré sphere as $N=1$. Based on this viewpoint, for a superposed mode consisting of $LG^p_{\pm\ell}(r,\varphi,z)$, the spatial amplitude is invariant upon paraxial propagation, given by $u^p_{|\ell|}(r,\varphi,z)e^{-ikz}[\sqrt{\alpha}e^{-i\ell\varphi} + e^{i\theta}\sqrt{1-\alpha}e^{i\ell\varphi}]$. Thus, we can reformulate Eq. (1) into

$$\mathbf{E}^\omega(r,\varphi,z) = u^p_{|\ell|}(r,\varphi,z)\exp[-ik(\omega)z]\hat{\mathbf{e}}_{\text{SOC}}$$
$$\hat{\mathbf{e}}_{\text{SOC}} = \sqrt{\alpha}\exp(-i\ell\varphi)\hat{\mathbf{e}}_+ + e^{i\theta}\sqrt{1-\alpha}\exp(i\ell\varphi)\hat{\mathbf{e}}_-, \quad (2)$$

where $k(\omega)$ is the wave vector, and $\hat{\mathbf{e}}_{\text{SOC}}$ denotes the SOC state describing spatially-variant SoPs.

The results mean that: first, the vector profiles obeying Eq. (2) are cylindrically symmetric and propagation invariant, and therefore, the corresponding vector light are known as cylindrical vector (CV) modes [43]; second, for CV modes, the profile of spatially-variant SoP, governed by $\hat{\mathbf{e}}_{\text{SOC}}$, is independent of the amplitude envelop $u^p_{|\ell|}(r,\varphi,z)$. In addition, for a given $\hat{\mathbf{e}}_\pm$ and the associated LG pair, all possible SOC states form a hybrid SU(2) space that can also be visualized as positions on an unit sphere, i.e., the so-called higher-order Poincaré sphere (HOPS) introduced by G. Milione *et al.* [44]. Notably, CV modes are interested for communication domain, because, first, they provide a set of propagation-invariant mutually unbiased bases (MUBs) for high-dimension quantum cryptography in free space [45, 46], while the scalar OAM version need imaging; and second, they are natural guiding modes of few-mode fiber, in contrast, OAM fiber is still developing.

The straightforward description for SPI upconversion of CV modes is: *to convert the frequency of arbitrary CV modes without changing its vector profile and position on the HOPS.* To realize this, we employed a SU(2) nonlinear interferometer based on the polarization Sagnac scheme with a type-II long crystal. Figure 1 shows a schematic of the apparatus, where the vector light is input from port-1 as the signal, the pump input from port-2, and the frequency converted signal (or SHG) is output from port-3. The input signal was first split using a dual-wavelength polarizing beam splitter (*d*-PBS), and if we define $\hat{\mathbf{e}}_+ = \sqrt{\beta}\hat{\mathbf{e}}_H + e^{i\phi}\sqrt{1-\beta}\hat{\mathbf{e}}_V$ and $\hat{\mathbf{e}}_- = \sqrt{1-\beta}\hat{\mathbf{e}}_H - e^{i\phi}\sqrt{\beta}\hat{\mathbf{e}}_V$, the signal in the interferometer can be represented as:

$$\mathbf{E}^\omega(r,\varphi,z) = E^\omega_H(r,\varphi,z)\hat{\mathbf{e}}_H + e^{i\phi}E^\omega_V(r,\varphi,z)\hat{\mathbf{e}}_V, \quad (3)$$

where $E^\omega_H(r,\varphi,z)$ and $E^\omega_V(r,\varphi,z)$ are two SoP-dependent spatial modes, with respect to $\hat{\mathbf{e}}_H$ and $\hat{\mathbf{e}}_V$, that propagate clockwise and anticlockwise in the Sagnac loop, respectively, see Appendix A for details. At port-2, a diagonal-polarized pump, given by $E^\omega_p(r,\varphi,z)\hat{\mathbf{e}}_D$, enters the loop with a transmission-reflection ratio of 50:50, where a Gaussian ($LG^0_0(r,\varphi,z)$) or flattop beam (i.e., super Gaussian beam) is employed as required as the pump. It should be noted that the Gaussian pump only works for the signal without a radial index (i.e., $p=0$), because the coupled spatial amplitude, $u_0^0(r,\varphi,z)*u_{|\ell|}^0(r,\varphi,z)$, is still a LG field but this does not hole for $p\neq 0$ [47, 48]. In the Sagnac loop, the presence of a dual-wavelength half wave plate (d-HWP) leads to a SoP swapping between $\hat{\mathbf{e}}_H$ and $\hat{\mathbf{e}}_V$ for any beams passing it, including pump, signal and SHG. In consequence, two type-II SHG processes [22], i.e., $E_H^\omega \hat{\mathbf{e}}_V * E_p^\omega \hat{\mathbf{e}}_H \to E_H^{2\omega}(r,\varphi,z)\hat{\mathbf{e}}_H$ and $E_V^\omega \hat{\mathbf{e}}_V * E_p^\omega \hat{\mathbf{e}}_H \to E_V^{2\omega}(r,\varphi,z)\hat{\mathbf{e}}_V$, occur for clockwise and anticlockwise trips of the signal beam, respectively. Then, the two generated scalar SHG with $\hat{\mathbf{e}}_H$ and $\hat{\mathbf{e}}_V$, respectively, overlap on the d-PBS again and output from a dichroic mirror (DM). Because of the phase-locking structure of the apparatus (because the fact of all beams propagating along the same loop), the intramodal phase $e^{i\phi}$ is maintained for the whole transformation. As a result, the SHG light appearing from port-3 is $\mathbf{E}^{2\omega}(r,\varphi,z) = E_H^{2\omega}(r,\varphi,z)\hat{\mathbf{e}}_H + e^{i\phi}E_V^{2\omega}(r,\varphi,z)\hat{\mathbf{e}}_V$, i.e., a SPI upconversion $\mathbf{E}^\omega(r,\varphi,z) \to \mathbf{E}^{2\omega}(r,\varphi,z)$ is achieved.

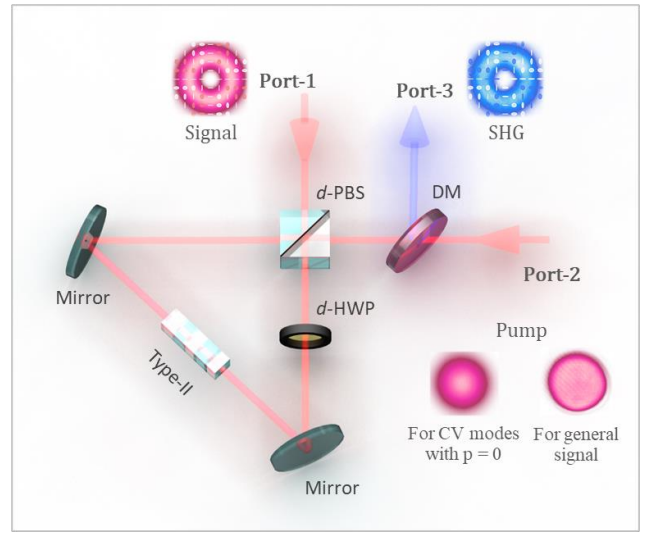

FIG. 1. Schematic of the polarization Sagnac nonlinear interferometer. The key components include the *dual-wavelength* polarizing beam splitter (*d*-PBS), *dual-wavelength* half-wave plate (*d*-HWP), and dichroic mirror (DM). The ellipses covered on the beam profiles depict spatial polarization distribution of vector light, where white and red (blue) represent right- and left-hand polarizations, respectively.

Figure 2 shows a schematic of the experimental setup, where a continuous laser at 800 nm (Toptica TA pro) was used as the fundamental frequency light for the SHG. The laser output from a single-mode fiber collimator was first converted into a perfect TEM$_{00}$ mode by passing through a spatial filter. Then, it was split using an HWP in combination with a polarizing beam splitter (PBS), where the transmitted and reflected parts were used for preparation of the signal and pump, respectively. For preparation of the signal, the reflected TEM$_{00}$ mode was first incident on a spatial light modulator (SLM-1, Holoeye PLUTO-2-NIR-080), where a phase hologram based on complex amplitude modulation was employed to extract the target LG mode. The extracted LG mode was then injected into a (two-arm) polarization Sagnac interferometer containing a Dove prism in one of the paths to transform it into the desired CV mode, and this served as the signal to be up converted. For preparation of the pump, the transmitted TEM$_{00}$ mode was sent to the SLM-2 and converted into a flattop or Gaussian beam with a variable size that depended on the requirements, and this served as the pump. For further details on the shaping light technique used see Ref. 49 and the MATLAB code in the reference. The prepared signal (~1 mW) and pump (~50 mW) were relayed into the port-1 and port-2 of the apparatus shown in Fig. 1, respectively. Then, they were focused into a bulk (10 mm long and 1×2 mm aperture) type-II periodically poled KTiOPO$_4$ (PPKTP) crystal using a 100 mm focal-length lens, and the crystal was placed in a temperature controller with stability of ±2 mK [50, 51]. At port-3, a dichroic mirror (DM) was used to separate the generated SHG (400 nm, 5~10 μW) from the residual pump.

To suppress the noise of pump SHG below the detectable threshold of the CCD (1 nW level for 400 nm), the o- and e-axes of the crystal were exactly aligned with the horizontal and vertical planes, respectively. Notice that, for signals at single-photon level, using frequency degenerate upconversion is necessary, such as the configuration reported in Refs. 52 and 53, so that the upconversion photons can be easily separated from the pump noise via frequency filtering. The inset in the bottom left of Fig. 2 shows the setup for the spatial Stokes measurement, which was used to characterize both the input signals and the corresponding SHG (see Ref. 54 for details about the characterization). In the whole optical setup, the reference frames of all polarization elements were exactly unified with a relation: p- and s-polarizations were corresponding to $\hat{\mathbf{e}}_H$ and $\hat{\mathbf{e}}_V$, respectively. Furthermore, both the signal preparation and SPI upconversion require high-quality PBS that can provide high polarization Extinction Ratio (PER) at both transmission and reflection ports. Specifically, the PER of the PBS used for signal preparation is: 3000:1 and 1000:1 for the transmission and reflection ports, respectively; and the PER of the $d$-PBS (800/400 nm) used for SPI upconversion is: 1000:1 and 500:1 for the transmission and reflection ports, respectively.

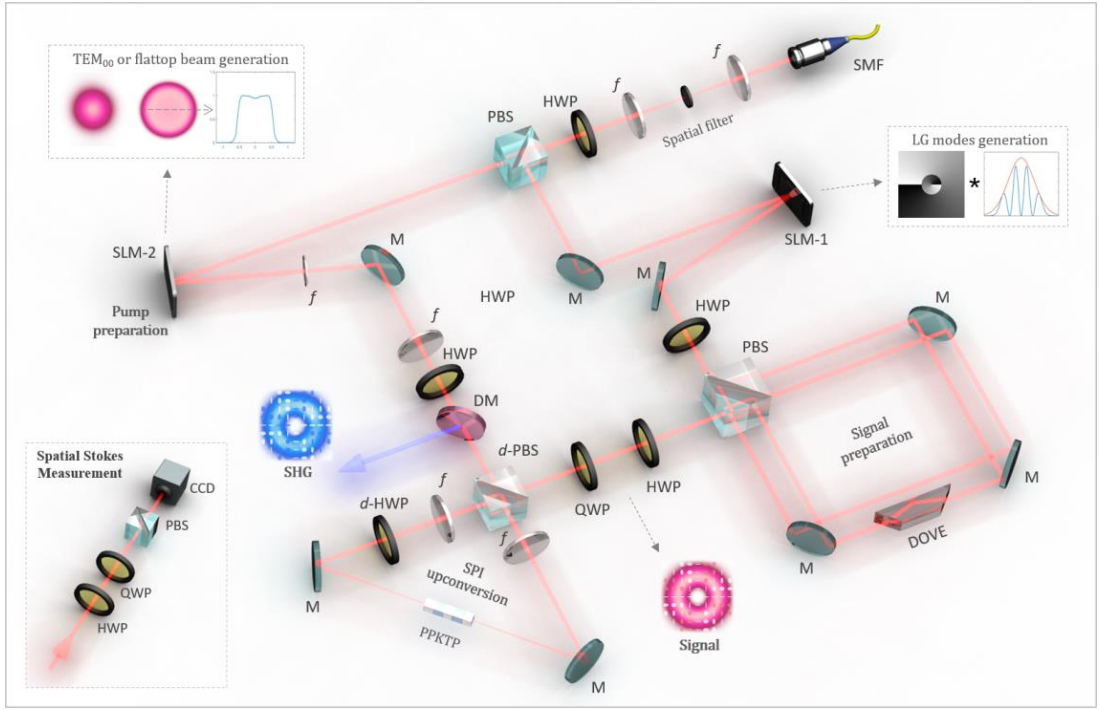

FIG. 2. Diagram of the experimental setup. The key components include the single mode fiber (SMF), polarizing beam splitter (PBS), half-wave plate (HWP), quarter-wave plate (QWP), Dove prism (DOVE), mirror (M), lens (*f*), dichroic mirror (DM), spatial light modulator (SLM), *dual-wavelength* polarizing beam splitter (d-PBS) and half-wave plate (d-HWP). The left bottom inset shows the setup for the spatial Stokes measurement.

To test conservation of the SOC states during the SPI upconversion, we first demonstrate the upconversion for all MUBs on a standard HOPS, introduced in Ref. 44, given by:

$$\begin{aligned}
\mathrm{I} &= \{|L_\ell\rangle = |\exp(-i\ell\varphi), \hat{\mathbf{e}}_L\rangle, |R_\ell\rangle = |\exp(i\ell\varphi), \hat{\mathbf{e}}_R\rangle\}; \\
\mathrm{II} &= \{|H_\ell\rangle = (|L_\ell\rangle + |R_\ell\rangle)/\sqrt{2}, |V_\ell\rangle = (|L_\ell\rangle - |R_\ell\rangle)/\sqrt{2}\}; \\
\mathrm{III} &= \{|D_\ell\rangle = (|H_\ell\rangle + |V_\ell\rangle)/\sqrt{2}, |A_\ell\rangle = (|H_\ell\rangle - |V_\ell\rangle)/\sqrt{2}\}.
\end{aligned} \quad (4)$$

More specifically, we choose the MUBs with $\ell = 1$, as shown in Fig. 3(a). We note that their vector profiles are rotationally invariant, and they are therefore an important resource for alignment-free quantum communication [14, 15]. Figures 3(b) and (c) show the vector profiles of the signal MUBs prepared experimentally and their corresponding upconversion, respectively. We see that the vector profiles of the two groups are in excellent agreement with each other, and with the theoretically expected results shown in Fig. 3(a). This great consistency of theory and observation originates from high-quality signal preparation and high-precision polarization-reference alignment. Specifically, first, exact LG modes were generated via complex-amplitude modulation, which were then converted into CV modes via the two-arm Sagnac interferometer. For clearly showing beam quality, the false-color beam profiles of signal and upconverted MUBs are provides in Appendix B. Based on the observed vector profiles of the signals and associated SHG output, their SOC states (or density matrices) can be accurately obtained according to the method in Ref. 54. This method, compared with the high-order Stokes tomography, can avoid the measuring error introduced by high-order diffraction noise of q-plates. Then, we can use the obtained states to plot their position on the HOPS, as red and blue points shown in Fig. 3(d). Additionally, we also performed complete MUB projections for them, i.e., projecting them onto the theoretical MUBs shown in Eq. (4), and the normalized outcomes (i.e., correlation matrices for the signals and SHG outputs) are shown in Figs. 3(e). These results verify SPI frequency conversion of the vector light with a rotationally invariant SOC structure. Moreover, an additional result for the MUBs with $\ell = 2$ is provided in Appendix B.

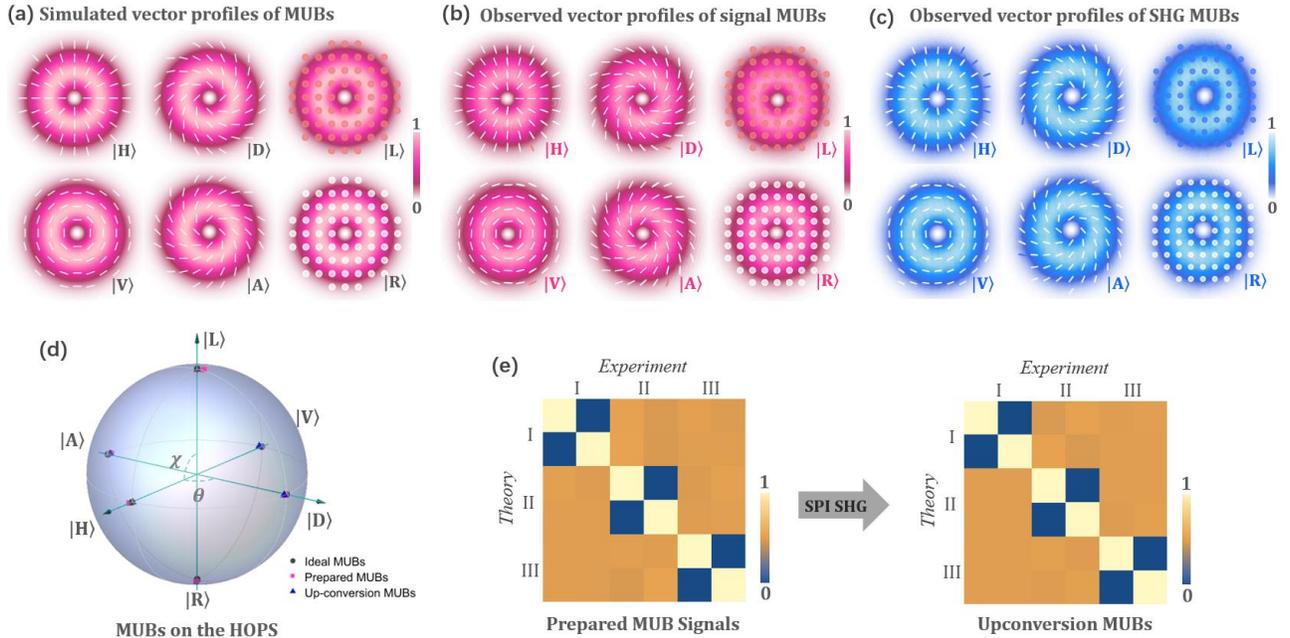

FIG. 3. SPI upconversion for complete MUBs of the HOPS spanned by $|L\rangle = |\hat{\mathbf{e}}_L, +1\rangle$ and $|R\rangle = |\hat{\mathbf{e}}_R, -1\rangle$, where (a)–(c) are vectorial profiles of the simulated MUBs, prepared signals and the corresponding SHG, respectively. (d) Positions of the MUBs on the HOPS. (e) Experimental correlation matrix for different MUBs obtained from the prepared signals and the corresponding SHG, respectively. The ellipses covered on the beam profiles depict the spatially-variant SoP, where white and red (blue) represent right- and left-hand polarizations, respectively.

Next, we demonstrate SPI upconversion of CV modes based on arbitrary SoP, whose vector profiles are no longer rotationally invariant. For this, we consider SOC states on two specific HOPSs, given by:

$$\sqrt{\alpha}\exp(-i\varphi)\hat{\mathbf{e}}_H + e^{i\theta}\sqrt{1-\alpha}\exp(i\varphi)\hat{\mathbf{e}}_V$$
$$\sqrt{\alpha}\exp(-i2\varphi)\hat{\mathbf{e}}_D + e^{i\theta}\sqrt{1-\alpha}\exp(i2\varphi)\hat{\mathbf{e}}_A, \quad (5)$$

where four states were chosen for each sphere, as shown by the gray points plotted on the HOPSs in Figs. 4(a) and (c). The experimental results, i.e., the positions of the experimentally prepared signals and the associated SHG output obtained via spatial Stokes measurement, are shown by the red and blue points, respectively, in Figs. 4(a) and (c). For both spheres, we see that each group of three points with different colors are in high proximity, indicating high accuracy for the signal preparation as well as the high fidelity of the upconversion. For a more intuitive display of the results, Figs. 4(b) and (d) show the vector profiles of the SOC states on the two spheres, including the theoretical expectations, observed signals, and corresponding SHG output. We see that all the observations for the signal and SHG are once again in excellent agreement with each other, as well as with the theoretically expected results. Additionally, a calculated SOC state fidelity based on the results is provided in Appendix B.

In above results, the spatial polarization independence of the upconversion was true for CV modes without the radial index, and all the SHG output were pumped by a Gaussian beam. From Eq. (2), it can be seen that to maintain the vector profile (or SOC structure) of a CV mode with $p=0$, using an easy-to-obtain Gaussian beam as the pump is sufficient. This is because the amplitude envelops of the NP excited by a beating field $u_0^0(r,\varphi,z) * u_{|\ell|}^0(r,\varphi,z)$ in the crystal still has an LG field. This particular result, however, is no longer true for the general cases, such as CV modes with $p \neq 0$, full-Poincaré modes [55] and polarization-resolved images. To demonstrate this, we consider a "radial polarized" signal with a well-defined radial index, i.e., $p=2$, given by $\sqrt{1/2}[LG_{+1}^2(r,\varphi,z)\hat{\mathbf{e}}_L + e^{i\theta}LG_{-1}^2(r,\varphi,z)\hat{\mathbf{e}}_R]$. Figures 5(a) and (b) show the upconversion of this signal pumped using a Gaussian beam. We see that, while its azimuthal-variant SoP is maintained in the upconversion, the well-defined radial structure was destroyed and is no longer propagation invariant. Moreover, our simulations have excellent agreement with the experimental observation; for this, see Appendix A for theoretical details, and in Ref. 48 we provide a general theory for the transformation of the radial mode of the LG beam in upconversion. To overcome this distortion, using a flattop (super Gaussian) beam as the pump is necessary [56], which can be easily obtained with high efficiency (~90% efficiency using a Holoeye NIR-080) via a phase-only light shaping technique [49]. Figures 5(c) and (d) show the upconversion pumped by a flattop beam, where we see that the radial amplitude of the SHG output agrees well with that of the signal for both near and far fields.

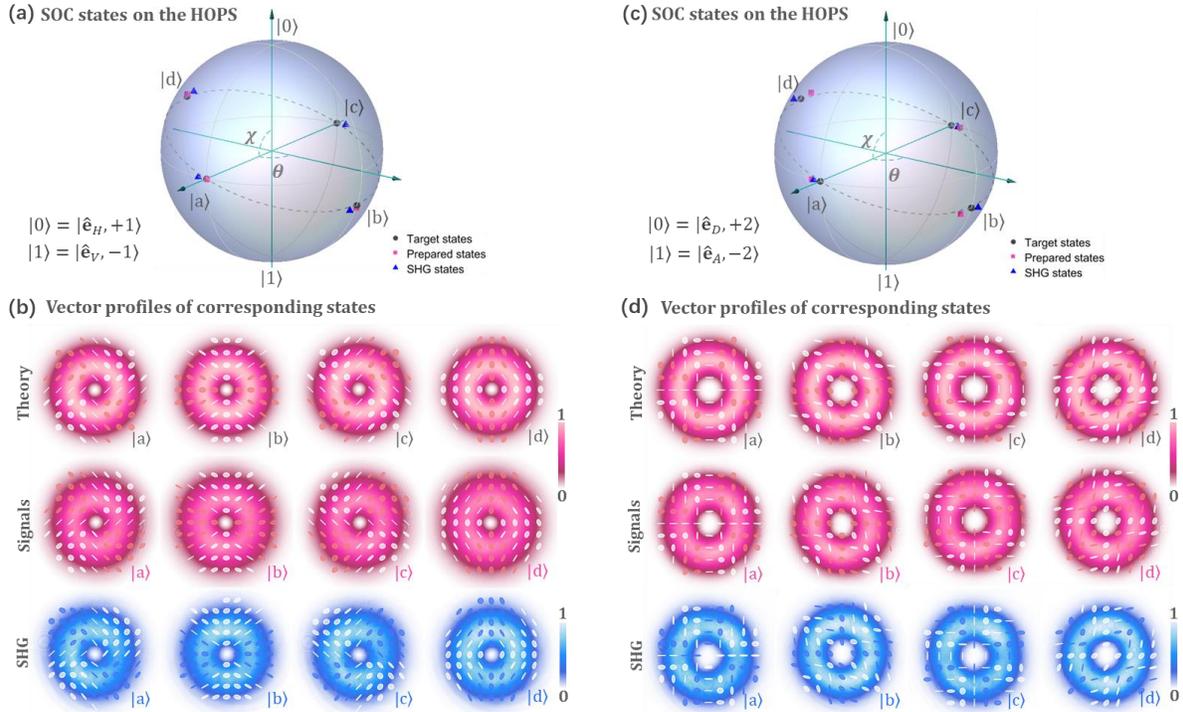

FIG. 4. SPI upconversion for the SOC states defined by an arbitrary SoP basis. (a) SOC states on the HOPS defined by $|0\rangle = |\hat{\mathbf{e}}_H, +1\rangle$ and $|1\rangle = |\hat{\mathbf{e}}_V, -1\rangle$, and (b) vector profiles of the corresponding states. (c) SOC states on the HOPS defined by $|0\rangle = |\hat{\mathbf{e}}_D, +2\rangle$ and $|1\rangle = |\hat{\mathbf{e}}_A, -2\rangle$, and (d) vector profiles of the corresponding states. The ellipses covered on beam profiles depict the spatially-variant SoP, where white and red (blue) represent right- and left-hand polarizations, respectively.

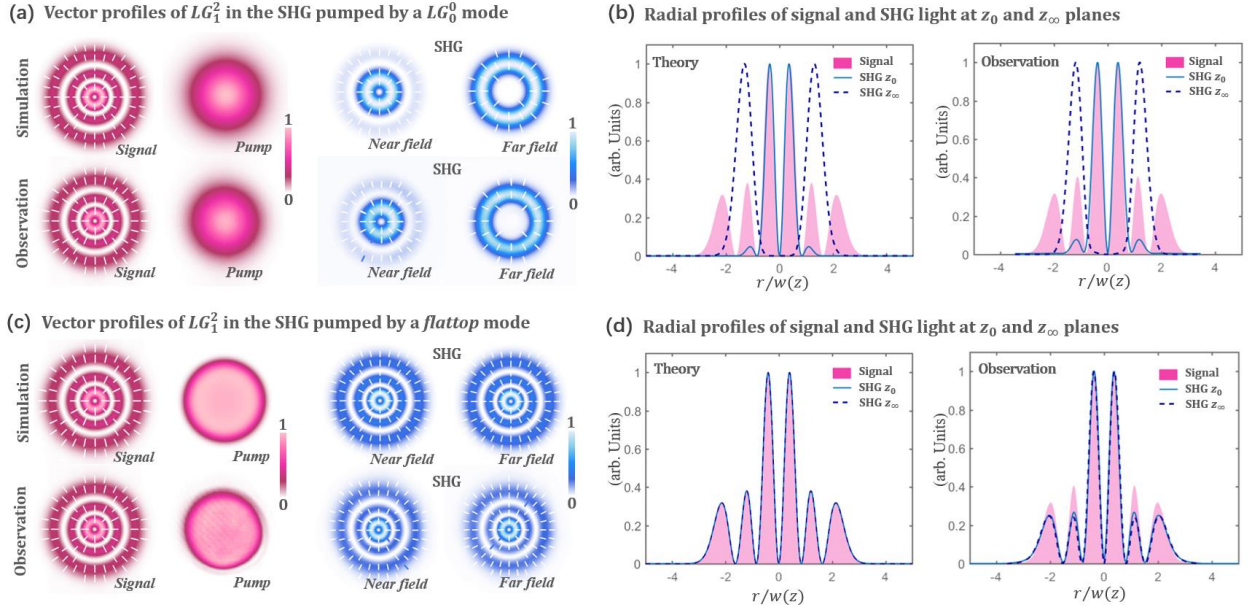

FIG. 5. SPI upconversion for the SOC states carrying radial structures. (a) Simulated and observed vector profiles of the SHG pumped by a Gaussian beam and (b) comparison of the corresponding radial profiles obtained by simulation and observation. (c) Simulated and observed vector profiles of the SHG pumped by a flattop beam, and (d) comparison of the corresponding radial profiles obtained by simulation and observation. In (b) and (d), to compensate the divergence, the x-axis unit is normalized about the beam waist $w(z)$, i.e., $r/w(z)$.

*Discussion and conclusion.* — SPI upconversion can benefit many aspects of optics field, especially for communication and imaging areas. First, it should be noted that, for both areas, frequency degenerate upconversion is a preferable configuration, because it is easier to operate and can avoid parametric noise (such as pump's SHG, SPDC and SRS). But one need to prepare high-quality triple-wavelength polarization elements, e.g., triple-wavelength PBS and waveplates. For communication, if the signal is carried by CV beams or photons (with $p=0$) [14–16, 45, 46], using an easy-to-obtain Gaussian beam or pulse as pump is enough; if the signal is a more general vector mode involving full-field spatial modes [57, 58], using a flattop pump is necessary. For imaging, using SPI upconversion can detect polarization-resolved images in Mid-/far-infrared region, and even enable time-polarization-resolved imaging when using ultrashort pulses as pump. Furthermore, within this context, flattop and Gaussian pumps provide different functions depending on the position (Fourier or intermediate plane) of the crystal [59, 60].

Additionally, the inverse process of SPI upconversion, i.e., PDC, also has many potential values. For instance, in the stimulated region, it enables frequency down conversion or phase conjugation for vector light [36]. In the spontaneous region, because the Sagnac nonlinear interferometer is an ideal apparatus to observe two-photon interference, thus a vector interference [61, 62] of two-photon version would happen when a solo vector pump input from the port-3 of the apparatus, and we will further demonstrate this in the future.

In summary, in this proof-of-principle work, we demonstrated SPI upconversion of vector light. The apparatus was based on a polarization Sagnac nonlinear interferometer with a type-II PPKTP crystal, where a flattop beam (or Gaussian beam in particular cases) with a variable beam size was employed as the pump. Our results show that the vector profile and the associated SOC state of the signal beam can be safely transferred into the upconversion beam with a high fidelity. The principle demonstrated here lays the foundation of SPI frequency interface that can used for high-dimension quantum or high-capacity classical channels based on vector modes [52, 53], and also pave the way for upconversion detection of polarization-resolved imaging [56, 59].

### ACKNOWLEDGMENT

This work was supported by the National Natural Science Foundation of China (Grant Nos. 11934013 and 61975047).

### Appendix A

According to the relation $\hat{\mathbf{e}}_+ = \sqrt{\beta}\hat{\mathbf{e}}_H + e^{i\phi}\sqrt{1-\beta}\hat{\mathbf{e}}_V$ and $\hat{\mathbf{e}}_- = \sqrt{1-\beta}\hat{\mathbf{e}}_H - e^{i\phi}\sqrt{\beta}\hat{\mathbf{e}}_V$, two SoP-dependent spatial modes in Eq. (3) is given by

$$E_H^\omega(r,\varphi,z) = u_\ell^p(r,\varphi,z)\exp[-ik(\omega)z]$$
$$\times \left[\sqrt{\alpha\beta}\exp(-i\ell\varphi) + e^{i\theta}\sqrt{(1-\alpha)(1-\beta)}\exp(i\ell\varphi)\right]$$
$$E_V^\omega(r,\varphi,z) = u_\ell^p(r,\varphi,z)\exp[-ik(\omega)z]$$
$$\times \left[\sqrt{(1-\alpha)\beta}\exp(-i\ell\varphi) - e^{i\theta}\sqrt{\alpha(1-\beta)}\exp(i\ell\varphi)\right]. \quad (6)$$

Now we consider the special case of $p=0$. For a Gaussian pump, i.e., $LG_0^0(r,\varphi,z) = u_0^0(r,\varphi,z)\exp[-ik(\omega)z]$, the excited NP of SHG in the clockwise and anticlockwise directions are given by

$$\mathbf{P}^{NL}_{clock.} \propto LG_0^0(r,\varphi,z_0)E_H^\omega(r,\varphi,z_0)$$
$$= u_0^0(r,\varphi,z_0)u_\ell^0(r,\varphi,z_0)\exp[-ik(2\omega)z]$$
$$\times \left[\sqrt{\alpha\beta}\exp(-i\ell\varphi)+e^{i\theta}\sqrt{(1-\alpha)(1-\beta)}\exp(i\ell\varphi)\right]$$
$$\mathbf{P}^{NL}_{anticlock.} \propto LG_0^0(r,\varphi,z_0)E_V^\omega(r,\varphi,z_0)$$
$$= u_0^0(r,\varphi,z_0)u_\ell^0(r,\varphi,z_0)\exp[-ik(2\omega)z]$$
$$\times \left[\sqrt{(1-\alpha)\beta}\exp(-i\ell\varphi)-e^{i\theta}\sqrt{\alpha(1-\beta)}\exp(i\ell\varphi)\right]. \quad (7)$$

Note, due to $u_0^0(r,\varphi,z_0)\big|_{|\ell|}^0(r,\varphi,z_0) \propto u_{|\ell|}^0(r,\varphi,z_0)$, thus we have $\mathbf{P}^{NL}_{clock.} \propto E_H^{2\omega}(r,\varphi,z)$ and $\mathbf{P}^{NL}_{anticlock.} \propto E_V^{2\omega}(r,\varphi,z)$, indicating a transformation of $\mathbf{E}^\omega(r,\varphi,z) \to \mathbf{E}^{2\omega}(r,\varphi,z)$ is achieved.

For the more general case of $p\neq 0$, notably, the excited NP $u_0^0(r,\varphi,z_0)*u_{|\ell|}^p(r,\varphi,z_0)$ is no longer an eigen LG field, but a superposed LG mode that has a well-defined azimuthal index and degenerated radial indices. More specifically, assuming a signal $LG_{|\ell|}^2(r,\varphi,z;w_0)$ is pumped by $LG_0^0(r,\varphi,z;nw_0)$, where $w_0$ and $nw_0$ are beam waist, according to Ref. 48, we have

$$\mathbf{P}^{NL} \propto LG_0^0(r,\varphi;nw_0)LG_{|\ell|}^2(r,\varphi;w_0)$$
$$= a_1 LG_{|\ell|}^0(r,\varphi;w')+a_2 LG_{|\ell|}^1(r,\varphi;w')+a_3 LG_{|\ell|}^2(r,\varphi;w'). \quad (8)$$

where $w'=nw_0/\sqrt{1+n^2}$ is the original beam waist of upconversion field and $a_{1,2,3}$ are modal weights, given by

$$a_1 = \sqrt{3}/\sqrt{3+6n^4+n^8}$$
$$a_2 = \sqrt{6}n^2/\sqrt{3+6n^4+n^8}$$
$$a_3 = n^4/\sqrt{3+6n^4+n^8} \quad . \quad (9)$$

From Eq. (8) and (9) we see that, first, the upconversion field is radial-index degenerated mode and the modal weights depends on the pump waist $nw_0$; second, $a_1,a_2$ decrease rapidly with the $n$ and become zero as $n\to\infty$ (i.e., a flattop beam), and Fig. 6 shows the functions $a_{1,2,3}^2(n)$. We now further consider the special case $n=1$, we have $a_1=\sqrt{3/10}$, $a_2=\sqrt{3/5}$ and $a_3=\sqrt{1/10}$. Obviously, due to containing different spatial-mode order $2p+|\ell|$ [41], the profile of upconversion field is not propagation invariant, and Fig. 5(a) shows the simulated vector beam profile at the near and far fields.

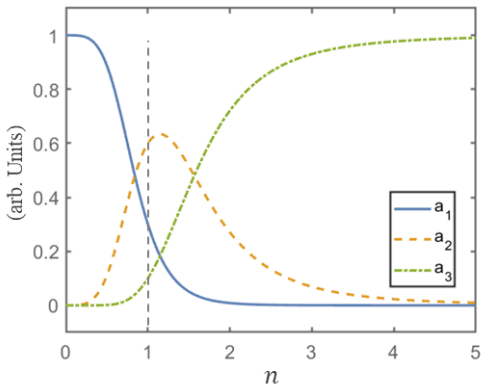

FIG. 6. Theoretical radial-mode spectra of the upconversion field as functions of $n$.

## Appendix B

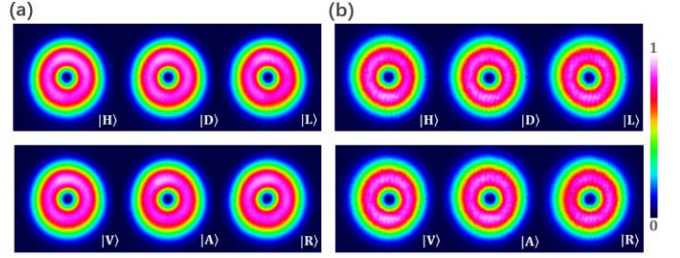

FIG. 7. False-color beam profiles of signal and upconverted MUBs, where (a) and (b) correspond to data in Fig. 3 (b) and (c), respectively.

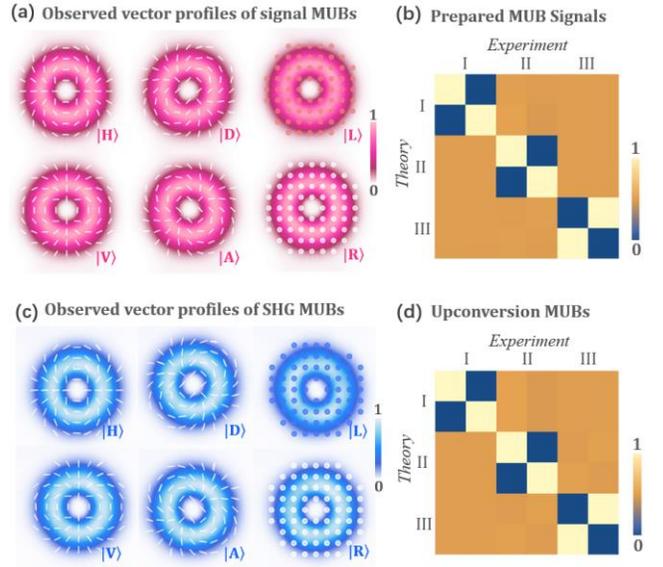

FIG. 8. Additional results for complete MUBs of the HOPS spanned by $|L\rangle = |\hat{\mathbf{e}}_L,+2\rangle$ and $|R\rangle = |\hat{\mathbf{e}}_R,-1\rangle$, where (a) and (b) are observed vectorial profiles of the signals and the corresponding correlation matrix, respectively; (c) and (d) are observed vectorial profiles of the SHG and the corresponding MUB correlation matrix, respectively.

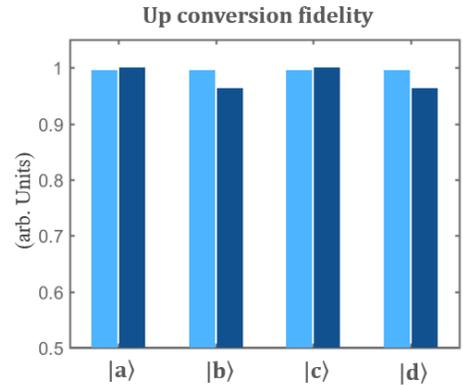

FIG. 9. Upconversion fidelity obtained from the inner products of the signals and their corresponding SHG, where light and dark blue corresponds to data in Fig. 4 (a) and (b), respectively.

# References

1. J. Chen, C.-H Wan, and Q.-W. Zhan. Sci. Bull. **63**, 54 (2018).
2. C. Rosales-Guzmán, B. Ndagano and A. Forbes. J. Opt. **20**, 123001 (2018).
3. H. Rubinsztein-Dunlop, *et al*. J. Opt. **19**, 013001 (2017).
4. K. Y. Bliokh, F. J. Rodriguez-Fortuno, F. Nori, *et al.* Nat. Photon. **9**, 796, (2015).
5. F. Cardano, L. Marrucci. Nat. Photon. **9**, 776 (2015).
6. C. V. S. Borges, M. Hor-Meyll, J. A. O. Huguenin, and A. Z. Khoury. Phys. Rev. A **82**, 033833 (2010).
7. Töppel, F., Aiello, A., Marquardt, C., Giacobino, E. & Leuchs, G. New. J. Phys. **16**, 073019 (2014).
8. D. Pohl. Appl. Phys. Lett. **20**, 266–267 (1972).
9. Y. Mushiake, K. Matzumurra, and N. Nakajima. Proc. IEEE **60**, 1107 (1972).
10. S. Quabis, R. Dorn, M. Eberler, O. Glöckl, and G. Leuchs. Opt. Commun. **179**, 1–7 (2000).
11. R. Dorn, S. Quabis, and G. Leuchs. Phys. Rev. Lett. **91**, 233901 (2003).
12. P. Török and P. Munro, Opt. Express **12**, 3605–3617 (2004).
13. R. Chen, K. Agarwal, C. J. R. Sheppard, and X. Chen. Optics lett. **38**, 3111-4 (2013).
14. V. D'Ambrosio, et al. Nat. Commun. **3**:961 (2012).
15. G. Vallone, et al. Phys.Rev. Lett. 113, 060503 (2014).
16. V. D'Ambrosio, et al. Phys. Rev. A **94**, 030304(R) (2016).
17. S. Berg-Johansen, F. Töppel, B. Stiller, P. Banzer, M. Ornigotti, E. Giacobino, G. Leuchs, A. Aiello, and C. Marquardt, Optica **2**, 864-868 (2015).
18. S. Slussarenko, A. Alberucci, C. P. Jisha, B. Piccirillo, E. Santamato, G. Assanto, L. Marrucci. Nat. Photon. **10**, 571-575 (2016).
19. L. Marrucci, E. Karimi, S. Slussarenko et al., J. Opt. **13**, 064001 (2011).
20. P. Chen, et al. Adv. Mater. 1903665 (2019)
21. L. Zhang, X. Qiu, F. Li, H. Liu, X. Chen, and L. Chen. Opt. Express, **26**(9), 11678 (2018).
22. H.-J. Wu, H.-R. Yang, C. Rosales-Guzmán, W. Gao, B.-S. Shi, and Z.-H. Zhu. Phys. Rev. A **100**, 053840 (2019).
23. L. J. Pereira, W. T. Buono, D. S. Tasca, K. Dechoum, and A. Z. Khoury. Phys. Rev. A **96**, 053856 (2017).
24. Stanislovaitis, P., Matijošius, A., Šetkus, V., & Smilgevičius, V. Lithuanian Journal of Physics, **54**(3) (2014).
25. Gu, B., Cao, X., Rui, G., & Cui, Y. Journal of Nonlinear Optical Physics & Materials, **27**(04), 1850045 (2018).
26. Y. Kozawa, and S. Sato. J. Opt. Soc. Am. B. 25(2), 175-179 (2008).
27. J. Vieira, R. M. G. M. Trines, E. P. Alves, R. A. Fonseca, J. T. Mendonça, R. Bingham, P. Norreys, and L. O. Silva. Nat. Commun. **7**, 10371 (2016).
28. Bouchard, F., Larocque, H., Yao, A. M., Travis, C., DeLeon, I., Rubano, A., Karimi, E., Oppo, G. L., Boyd, R. W. Phys. Rev. Lett. **117**, 233903 (2016).
29. Z.-H. Zhu, P. Chen, L.-W. Sheng, Y.-L. Wang, W. Hu, Y.-Q. Lu, and W. Gao, Appl. Phys. Lett. **110**, 141104 (2017).
30. F. Kong, C. Zhang, H. Larocque, F. Bouchard, Z. Li, M. Taucer, G. Brown, S. Severino, T. J. Hammond, E. Karimi, and P. B. Corkum. Phys. Rev. Research **1**(3), 032008, (2019).
31. F. Kong, H. Larocque, E. Karimi, P.B. Corkum, C. Zhang. Optica **6**(2), 160-164, (2019).
32. M. V. Chekhova and Z. Y. Ou, Adv. Opt. Photon. **8**, 104 (2016)
33. P. G. Kwiat, E. Waks, A. G. White, I. Appelbaum, and P. H. Eberhard. Phys. Rev. A **60**, R773 (1999).
34. B.-S. Shi and A. Tomita. Phys. Rev. A **69**, 013803 (2004).
35. T. Kim, M. Fiorentino, and F. N. C. Wong. Phys. Rev. A **73**, 012316 (2006).
36. André G. de Oliveira, Marcelo F. Z. Arruda, Willamys C. Soares, Stephen P. Walborn, Rafael M. Gomes, Renné Medeiros de Araújo, and Paulo H. Souto Ribeiro. ACS Photonics 7, 249–255 (2020)
37. H. Liu, H. Li, Y. Zheng, and X. Chen, Opt. Lett. **43**, 5981-5984 (2018).
38. R. K. Saripalli, A. Ghosh, A. Chaitanya, and G. K. Samanta. Appl. Phys. Lett. **115**, 051101 (2019).
39. H.-J. Wu, Z.-Y. Zhou, W. Gao, B.-S. Shi, and Z.-H. Zhu. Phys. Rev. A **99**, 023830 (2019).
40. C. Yang, Z.-Y. Z HOU Y. Li, Y.-H. Li, S.-L. Liu, S.-K. Liu, Z.-H. Xu, G.-C. Guo, and B.-S. Shi. Opt. Lett. **44**, 219-222 (2019).
41. L. Allen, J. Courtial, and M. J. Padgett. Phys. Rev. E **60**, 7497 (1999).
42. M. J. Padgett and J. Courtial, Opt. Lett. **24**, 430 (1998).
43. Zhan Q. Adv. Opt. Photon. **1**,1 (2009).
44. G. Milione, H. I. Sztul, D. A. Nolan, and R. R. Alfano, Phys. Rev. Lett. **107**, 053601 (2011).
45. E. Nagali, L. Sansoni, L. Marrucci, E. Santamato, and F. Sciarrino. Phys. Rev. A **81**, 052317 (2010).
46. A. Sit, F. Bouchard, R. Fickler, J. G.-Bischoff, H. Larocque, D. Heshami, D. Elser, C. Peuntinger, K. Günthner, B. Heim, C. Marquardt, G. Leuchs, R. W. Boyd, and E. Karimi. Optica **4**(9), 1006-1010 (2017).
47. J. Courtial, K. Dholakia, L. Allen, and M. J. Padgett. Phys. Rev. A **56**, 4193 (1997).
48. H.-J. Wu, L.-W. Mao, Y.-J. Yang, C. Rosales-Guzmán, W. Gao, B.-S. Shi, Z.-H. Zhu. arXiv:1912.05585.
49. C. Rosales-Guzmán, and A. Forbes. *How to Shape Light with Spatial Light Modulators.* (SPIE Press Book, 2017).
50. Z.-H. Xu, Y.-H. Li, Z.-Y. Zhou, S.-L. Liu, Y. Li, S.-K. Liu, C. Yang, G.-C. Guo, and B.-S. Shi. Opt. Express **28**, 5077-5084 (2020).
51. Y. Li, Z.-Y. Zhou, D.-S. Ding, and B.-S. Shi. Chin. Opt. Lett. **12**, 111901- (2014).
52. Z.-Y. Zhou, S.-L. Liu, Yan Li, D.-S. Ding, W. Zhang, S. Shi, M.-X. Dong, B.-S. Shi, and G.-C. Guo. Phys. Rev. Lett. 117, 103601 (2016).
53. S.-L. Liu, C. Yang, Z.-H. Xu, *et al.* Phys. Rev. A 101, 012339 (2020).
54. B.-S. Yu, H.-J. Wu, H.-R. Yang, W. Gao, C. Rosales-Guzman, B.-S. Shi, Z.-H. Zhu. arXiv:1907.04035.
55. A. M. Beckley, T. G. Brown, and M. A. Alonso, Opt. Express **18**, 10777 (2010).
56. J. S. Dam, C. Pedersen, and P. Tidemand-Lichtenberg. Opt. Express **20**(2), 1475-1482 (2012)
57. V. D. Salakhutdinov, E. R. Eliel, and W. Löffler. Phys. Rev. Lett. **108**, 173604 (2012).
58. Hiekkamäki, Markus; Prabhakar, Shashi; Fickler, Robert. Opt. Express **27**(22) 31456 (2019).
59. A. Barh, P. J. Rodrigo, L. Meng, C Pedersen, and P. Tidemand-Lichtenberg. Adv. Opt. Photon., **11**(4): 952-1019 (2019).
60. X. Qiu, F. Li, W. Zhang, Z. Zhu, and L. Chen. Optica **5**(2), 208-212 (2018).
61. A. Norrman, A. T. Friberg, and G. Leuchs. Optica **7**, 93-97 (2020).
62. E. Otte, C. Rosales-Guzmán, B. Ndagano, C. Denz, and A. Forbes. Light Sci Appl **7**, 18009 (2018).